\documentstyle[12pt]{article}
\begin{document}
\def \beq{\begin{equation}}
\def \eeq{\end{equation}}
\def \g{{\rm GeV}}
\def \msbar{\overline{\rm MS}}
\def \s{\hat s^2}
\def \seff{\sin^2 \theta_{\rm eff}^{\rm lept}}
\def \st{\sqrt{2}}
\rightline{EFI-99-07}
\rightline{hep-ph/9903219}
\rightline{March 1999}
\vspace{0.3in}
\centerline{\bf PARTICLES IN LOOPS -- FROM ELECTRONS TO TOP QUARKS
\footnote{Dedicated to the memory of Professor Hiroshi Suura.  Based on
a colloquium in his honor at the University of Minnesota, 1 June 1994,
updated to February 1999.}}
\vspace{0.3in}
\centerline{\it Jonathan L. Rosner}
\centerline{\it Enrico Fermi Institute and Department of Physics}
\centerline{\it University of Chicago, Chicago, IL 60637}
\bigskip

\centerline{\bf ABSTRACT}
\medskip
\begin{quote}

This article, in memory of Professor Hiroshi Suura, is devoted to the
effects of particles in loops, ranging from quantum electrodynamics to
precise tests of the electroweak theory and CP violation.
\end{quote}

\centerline{\bf INTRODUCTION}
\bigskip

I owe an enormous debt to Hiroshi Suura.  It was partly work \cite{Z3} on the
subject of this article that led him to bring me to the University of
Minnesota, where I spent 13 pleasant years.  He was an early collaborator
\cite{RS}, teaching me the value of clear thinking and careful statements.
Throughout the years, he was a constant source of good ideas, sound judgement,
and friendly advice.  He was responsible for the contacts that led to my first
visit to Japan in 1973, during which the generous hospitality my family and I
received led us to return time and again to a country for which we have great
love and admiration.  During one such visit in 1981 I was privileged to meet
Hiroshi's sister and brother-and-law on the occasion of a Japanese Physical
Society meeting in Hiroshima at the end of March.  I am thus especially honored
to be able to pay tribute to Hiroshi's memory for a similar meeting eighteen
years later.  I miss him greatly. 

I shall not discuss Hiroshi's important contributions to the theory of infrared
corrections \cite{IR}.  This work has been central to a wide variety of
experiments in elementary particle physics, particularly those involving
electrons.  Many of the precise measurements I shall describe could not have
been done without it.  However, another theme running through Hiroshi's work
and connecting it to the major issues of today's particle physics is the idea
of ``particles in loops.'' One of his most-quoted results concerns the effect
of electron loops in the calculation of the muon's anomalous magnetic moment
$a_\mu$ \cite{SWP}. This leads to a difference between $a_\mu$ and the
corresponding quantity $a_e$ for the electron, which was confirmed in beautiful
experiments at CERN \cite{firstmu} and is still the subject of intense scrutiny
\cite{BNLgmu}. Hiroshi once admitted his reluctance to be known for a
calculation which took him such a short time.  But his key contribution was not
only in performing the calculation, but in being able to do so and in knowing
what calculation to perform. 

The effects of ``particles in loops'' indeed permeate almost all of today's
high energy physics.  They have allowed us to make fundamental discoveries
about the properties of quarks, to anticipate the charmed quark's existence and
the top quark's mass, and to understand, at least in part, the violation of CP
symmetry.  This article briefly reviews those effects. For more technical
details (some of which will be updated here) see, e.g., Ref.~\cite{cmts}.  Many
of the historical references are taken from \cite{FR}. 

In Section II we discuss vacuum polarization and radiative corrections. Section
III is devoted to specific effects of quarks and leptons in loops. We review
electroweak unification in Section IV, and CP violation in Section V. Some
speculations on composite Higgs bosons and composite fermions occupy Sections
VI and VII, respectively. Section VIII summarizes. 
\bigskip

\centerline{\bf II.  VACUUM POLARIZATION AND RADIATIVE CORRECTIONS}
\bigskip

\leftline{\bf A.  Vacuum polarization}
\bigskip

The large positive charge of a nucleus ``polarizes the vacuum.'' Virtual
electrons are attracted to the nucleus, while virtual positrons are repelled. A
test charge at large distances sees the nucleus screened by the electrons,
while at short distances it penetrates the screening cloud and sees a larger
charge.  In quantum electrodynamics this may be thought of as the effect of an
electron-positron ``loop'' in the photon propagator.  A direct calculation of
this effect finds it to be infinite!  However \cite{Schw}, one can circumvent
this difficulty by comparing {\it ratios} of effective charges at two different
distance scales. Defining the fine-structure constant $\alpha \equiv e^2/4 \pi
\hbar c$ in terms of the charge $e$, and momentum scales $\mu_i = \hbar/r_i$ in
terms of distance scales $r_i~(i=1,2)$, the lowest-order result is 
\beq \label{eqn:lan}
\alpha(\mu_1) = \frac{\alpha(\mu_2)}{1 - (\alpha/3 \pi) \ln
(\mu_1^2/\mu_2^2)}
\eeq
and hence $\alpha(\mu_1) > \alpha(\mu_2)$ for $\mu_1 > \mu_2$.  The
electromagnetic interaction thus becomes stronger at higher momentum scales
(shorter distance scales).  For an electron bound in hydrogen \cite{SU}, vacuum
polarization leads to a stronger attraction in a $2S$ state than in a $2P$
state, leading to a splitting between the levels of $\Delta E(2S-2P) = - 27$
MHz.
\bigskip

\leftline{\bf B.  The Lamb shift}
\bigskip

The experimental value of the $2S-2P$ splitting in hydrogen was first measured
by W. Lamb in 1947 \cite{Lamb}.  In addition to the vacuum polarization effect
mentioned above, there is a much more substantial shift in the other direction,
in which an electron emits and reabsorbs a virtual photon while interacting
with the nucleus \cite{Lambth}.  The most recent experimental values for the
splitting are $1057.8514 \pm 0.0019$ MHz \cite{Palch}, $1057.845 \pm 0.009$ MHz
\cite{LP}, and $1057.839 \pm 0.012$ MHz \cite{HP}, to be compared with the
theoretical calculation \cite{Pach} of $1057.838 \pm 0.006$ MHz. 
\bigskip 

\leftline{\bf C.  The electron $g$-factor}
\bigskip

The process in which an electron emits and reabsorbs a virtual photon while
interacting with an external field also alters its magnetic moment $\mu_e$,
expressed in terms of its spin $\vec{S_e}$ via a quantity $g$: $\vec{\mu_e} =
\vec{S_e} g e /(2 m_e c)$.  In the Dirac theory of the electron, $g=2$.  The
correction to this result is \cite{Kin,Kinnew}
\beq \label{eqn:ge}
\left. \frac{g-2}{2} \right |_e = \frac{\alpha}{2 \pi} - 0.328~476~965
\left( \frac{\alpha}{\pi} \right)^2 + \ldots = (1~159~652~140 \pm
27) \times 10^{-12}~~~.
\eeq
The lowest-order term is due to Schwinger \cite{Schwg}; the corrections have
been calculated up to ${\cal O}(\alpha^5)$. The error is due mainly to
uncertainty in $\alpha$.  The latest experimental result \cite{Dehmelt} is 
\beq
\left. \frac{g-2}{2} \right |_e = (1~159~652~188 \pm 3) \times 10^{-12}~~~.
\eeq
The agreement with theory, and that of the Lamb shift mentioned earlier, are
examples of the successful application of quantum field theory to
electrodynamics. 
\bigskip

\leftline{\bf D.  The muon $g$-factor}
\bigskip

The second term on the right-hand side of Eq.~(\ref{eqn:ge}), $-0.328 \ldots
(\alpha/\pi)^2$, contains a contribution from the process in which an electron
emits and reabsorbs a virtual photon while interacting with the external field,
and this virtual photon itself is subject to the vacuum polarization effect
(\ref{eqn:lan}).  The virtual photon thus can be affected by any charged
particle-antiparticle pair in a loop. The pair providing the major contribution
to the electron $g$-factor is an electron-positron pair; other heavier
particles contribute, but not significantly to this order. 

For the muon $g$-factor, the situation is different.  Here, both the $e$-loop
and the $\mu$-loop are important.  The major effect of the $e$-loop can be
regarded as an effective modification of the leading-order $\alpha/2 \pi$
correction:
\beq
\frac{\alpha}{2 \pi} \to \frac{\alpha}{2 \pi} \left[ 1 - \frac{\alpha}{3 \pi}
\left( \ln \frac{m_\mu^2}{m_e^2} - {\rm const}\right) \right]^{-1}~~~
\eeq
as dictated by the correction (\ref{eqn:lan}).  The fine-structure constant
``runs'' as a function of distance (i.e., momentum) scale.  The theoretical
expression for the muon $g$-factor thus differs from that for the electron at
second order in $\alpha$ \cite{SWP}.  This observation of Hiroshi's was a
shining example of how particles in loops {\it other than those under direct
study} can affect measurable physics.  We shall see a number of more recent
applications of this idea in subsequent sections. 

The present theoretical expression \cite{gmuth,Cz,Davier} for the muon
$g$-factor is 
\beq \label{eqn:gm}
\left. \frac{g-2}{2} \right |_\mu = \frac{\alpha}{2 \pi} +
0.765~857~388(44) \left( \frac{\alpha}{\pi} \right)^2 + \ldots = (11~659~159.6
\pm 6.7) \times 10^{-10}~~~, 
\eeq
to be compared with the experimental value \cite{gmuex} $(11~659~230 \pm 85)
\times 10^{-10}$.  At this level of accuracy one must consider the effects of
not only electrons and muons in loops, but also quarks. A new experiment at
Brookhaven National Laboratory seeks to probe $a_\mu$ 20 times more precisely,
reaching enough sensitivity to probe even the effect of weakly interacting
particles in loops \cite{BNLgmu}. 
\bigskip

\centerline{\bf III.  QUARKS AND LEPTONS IN LOOPS}
\bigskip

\leftline{\bf A.  Neutral pion decay}
\bigskip

The decay of the neutral pion $\pi^0$ is governed by a triangle ``anomaly''
diagram \cite{JS}.  The $\pi^0$ dissociates into a quark-antiquark pair which
then annihilates into two photons. The process thus counts the number of quarks
$q$ traveling around the loop, weighted by the product of their coupling to the
$\pi^0$ and the square of their charges $Q(q)$.  Since the $\pi^0$ is
represented in the quark model as $(u \bar u - d \bar d)/\sqrt{2}$, the
amplitude for $\pi^0 \to \gamma \gamma$ thus measures $S = \sum[Q(u)^2 -
Q(d)^2]$, where the sum is taken over the number of quark species (``colors'',
if one wishes).  For 3 colors of fractionally charged (``Gell-Mann--Zweig'')
\cite{GZ} quarks, $S = 3[(2/3)^2 - (-1/3)^2] = 1$. 

An alternative quark model involves integrally charged (``Han--Nambu'') quarks
\cite{HN}:  Two colors of $u$ quark have $Q(u_{1,2}) = 1$ while one color has
$Q(u_3) = 0$; two colors of $d$ quarks have $Q(d_{1,2}) = 0$ while one color
has $Q(d_3) = -1$.  The amplitude for $\pi^0 \to \gamma \gamma$ turns out to be
the same \cite{Okubo}. Hiroshi was intrigued with this possibility \cite{OS},
and we had many interesting discussions on the subject.  It is interesting that
quarks at high density may undergo a color-flavor ``locking'' which converts
them from the Gell-Mann--Zweig to the Han--Nambu variety \cite{Wilcz}.  It is
one of many results in the past year on which I would have enjoyed hearing
Hiroshi's opinion. 
\bigskip

\leftline{\bf B.  Triangle anomalies and fermion families}
\bigskip

The triangle anomaly's contribution to trilinear gauge boson couplings is
undesirable in unified theories of the weak and electromagnetic interactions.
In order that it vanish, the sum of $I_{3L}Q^2$ over all fermions must equal
zero.  Here $I_{3L}$ is ``left-handed isospin,'' equal to 1/2 for left-handed
$u$ quarks and neutrinos, $-1/2$ for left-handed $d$ quarks and charged leptons
$\ell^-$, and zero for all left-handed antiparticles. This sum vanishes for
quarks and leptons within a single ``family,'' with respective contributions of
2/3, $-1/6$, 0, and $-1/2$ from, e.g., $u,~d,~\nu_e,~e^-$. The need for the
charmed quark in the second family $c,~s,~\nu_\mu,\mu^-$ was in part argued
\cite{BIM} on the basis of anomaly cancellation. Definitive evidence for the
charmed quark was presented within two years \cite{charm}, in the form of a $c
\bar c$ bound state, the $J/\psi$ particle.  (There had already been
indications of charm in cosmic ray events \cite{Niu}, which were taken very
seriously in Japan \cite{charmth}.) 

The anomaly cancellation confirmed by the charmed quark's discovery was
short-lived.  A third lepton $\tau$ was announced within the year \cite{Perl}.
A third pair of quarks $t,~b$ (proposed earlier \cite{KM} to explain CP
violation; see Sec. V) was then required to restore the cancellation \cite{HH}.
The $b$ was discovered in 1977 \cite{ups} and the $t$ in 1994 \cite{top}, both
at Fermilab.  The high mass of the top quark, $m_t = 174 \pm 5$ GeV/$c^2$
\cite{topmass}, makes it a particularly important player in many loop diagrams,
in ways which we now describe. 
\bigskip

\centerline{\bf IV.  ELECTROWEAK UNIFICATION}
\bigskip

\leftline{\bf A.  The SU(2) $\times$ U(1) gauge theory}
\bigskip

Fifty years ago it was popular to talk of the ``four forces of Nature'':
gravity, electromagnetism, the weak force, and the strong force.  We sometimes
forget that Newton's theory of gravity itself was a unfication of terrestrial
and celestial phenomena, while Maxwell's theory of electromagnetism, building
upon Faraday's experiments, unified previously distinct electrostatic and
magnetic results.

During Hiroshi's career we have seen the successful unification of the weak and
electromagnetic interactions \cite{GWS}.  In analogy with the view of
electromagnetism as arising from photon exchange, we now view the weak
interactions (those responsible, for example, for nuclear beta-decay) as
arising from the exchange of charged, massive $W$ bosons.  The unified theory
allows self-consistent calculations of weak processes at high energies and to
higher orders of perturbation theory.  The prices to pay are that (1) the
$W^\pm$ must exist (it was discovered in 1983 \cite{Wdisc}), and (2) the
simplest version also requires a massive {\it neutral} boson, the $Z^0$ (also
discovered in 1983 \cite{Zdisc}).  The exchange of a $Z$ leads to new weak
charge-preserving interactions, first seen in 1973 \cite{NC}. 

The new theory has the symmetry SU(2) $\times$ U(1), broken to U(1) of
electromagnetism by the mechanism which gives the $W^\pm$ and $Z^0$ bosons
their masses while leaving the photon massless.  The neutral SU(2) boson,
$W^0$, and the U(1) boson, $B^0$, mix with an angle $\theta$ to give the
massless photon and the massive $Z^0$. 

Since quarks of the same charge can mix with one another, the charge-changing
transitions involving $W$ emission and absorption connect all quarks of charge
2/3 with all quarks of charge $-1/3$ through a unitary matrix $V$, the {\it
Cabibbo-Kobayashi-Maskawa} (CKM) \cite{KM,Cab} matrix.  As a result of the
unitarity of $V$, the couplings of $Z^0$ remain diagonal in quark ``flavor''
even after mixing.  The only corrections to the flavor-diagonal nature of
neutral weak processes come at higher orders of perturbation theory, through
particles in loops.
\bigskip

\leftline{\bf B.  Main electroweak corrections}
\bigskip

A major source of corrections to the electroweak theory, which can now be
probed as a result of the precision of varied experiments, is the effect of
particles in loops in the photon, $W$, and $Z$ propagators.

All charged fermions can contribute in pairs to the photon charge
renormalization (the effect of Eq.~(\ref{eqn:lan}) and its higher-order
generalizations).  Whereas at long distances the fine structure constant
$\alpha$ is approximately 1/137.036, when probed at the scale of the $Z^0$ mass
it is \cite{Davier} $\alpha(M_Z) \simeq 1/128.9$.  This simple correction
substantially improves the predictions of the unified theory for the $W$ and
$Z$ masses, given the value of the electroweak mixing parameter $\sin^2 \theta
= 0.23156 \pm 0.00019$ measured in a wide variety of neutral-current processes
\cite{Karlen}. 

The $W$ and $Z$ propagators receive large contributions from loops involving
the third quark family as a result of the large top quark mass.  The prediction
of the lowest-order electroweak theory, $M_W/M_Z = \cos \theta$, is modified to
\cite{Veltman} 
\beq
\frac{M_W^2}{M_Z^2} = \rho \cos^2 \theta~~,~~~
\rho \simeq 1 + \frac{3 G_F m_t^2}{8 \pi^2 \sqrt{2}}~~~.
\eeq
Here $G_F = 1.16639 \times 10^{-5}$ GeV$^{-2}$ is the Fermi coupling constant,
and $m_t = 174 \pm 5$ GeV/$c^2$ is the top quark mass.  The parameter $\rho$ is
then about a percent, and multiplies the amplitude of every weak
neutral-current process.  Consequently, each of these processes probes $m_t$,
so it was possible to anticipate its value (modulo effects of the Higgs boson,
which we discuss next) before it was measured directly. 
\bigskip

\leftline{\bf C.  The Higgs boson and its effects}
\bigskip

A consequence of endowing the $W$ bosons with mass is that the elastic
scattering of longitudinally polarized $W^+ W^-$ does not have acceptable
high-energy behavior.  It would violate the unitarity of the $S$-matrix (i.e.,
would violate probability conservation) at high energies unless a spinless
neutral boson (the ``Higgs boson'') exists below a mass of $M_H \simeq 1$
TeV/$c^2$ \cite{LQT}.  The discovery of such a boson is a prime motivation for
multi-TeV hadron colliders such as the Large Hadron Collider (LHC) now under
construction at CERN.  Searches in $e^+ e^-$ collisions at LEP find no evidence
for the Higgs boson below nearly 100 GeV/$c^2$ \cite{Karlen}, but precision
electroweak experiments seem to favor a Higgs mass near this lower limit. 

Virtual Higgs boson can contribute to loops in the $W$ and $Z$ propagators,
thus affecting not only $\rho$ but a parameter $S$ \cite{PT} which expresses
the difference between electroweak results at low momentum transfers and those
probed at the higher momentum scale of $Z^0$ decays. One can calculate all
electroweak observables for nominal values of $m_t$ and $M_H$ (say, 175 and 300
GeV$/c^2$, respectively) and then ask how they deviate from those nominal
values, thereby specifying constraints on the parameters $\rho$ and $S$.  Given
the observed value of $M_Z$, one obtains \cite{cmts} a nominal value of $\sin^2
\theta = 0.2321 \equiv x_0$. It is conventional to define $\Delta \rho =
\alpha T$, and one then finds \cite{PT} 
\beq
T \simeq \frac{3}{16 \pi x_0} \left[ \frac{m_t^2 - (175 {\rm~GeV})^2}
{M_W^2} \right] - \frac{3}{8 \pi (1-x_0)} \ln \frac{M_H}{300 {\rm~GeV}}~~~.
\eeq
Note the quadratic dependence on $m_t$, but only logarithmic dependence on
$M_H$.  That is why electroweak observables were able to predict a top quark
mass (with some uncertainty) despite the absence of information about the Higgs
boson mass. The ``$S$'' parameter is logarithmic in both $m_t$ and $M_H$. As
in the case of $T$, it can be defined to be zero for nominal values of $m_t$
and $M_H$, so that deviations of $S$ from zero are indicative of new physics.

Fits to a wide variety of electroweak parameters are performed periodically as
these data become more and more precise.  Such data include the ratio of
charge-preserving to charge-changing deep inelastic cross sections for
neutrinos on matter, the $W$ mass (measured at LEP and Fermilab), and a host
properties of the $Z$ boson, such as its mass, width, branching ratios, and
decay asymmetries (measured at LEP and the Stanford Linear Collider). Since the
Higgs boson appears in loops, and the top quark mass is fairly well pinned
down, such fits can constrain the (logarithm of the) Higgs boson mass. A recent
fit \cite{Karlen} finds $M_H = 84^{+91}_{-51}$ GeV/$c^2$ \cite{mod}, or $M_H <
280$ GeV at 95\% confidence level.  Of course, much of this range is already
ruled out by the direct searches mentioned earlier. 
\bigskip

\leftline{\bf D.  Effects of other new particles; atomic parity violation}
\bigskip

The $S$ and $T$ parameters respond differently to new particles.  The $T$
parameter is affected by the presence of pairs of left-handed fermions with
charges differing by one unit (such as $t$ and $b$) whose masses also differ
from one another (as in the case of $t$ and $b$).  However, it is not affected
by new degenerate pairs.  The $S$ parameter, on the other hand, {\it is}
affected.  It is a good probe of new particles in loops, even if these
particles hide their contributions to $T$ by being degenerate in mass, and no
matter how heavy these particles may be \cite{PT}. 

One probe of $S$ is almost insensitive to $T$ \cite{MR,Sandars}.  Atomic
transitions can violate mirror symmetry (parity) as a result of the
interference of photon and $Z$ exchange. The coherent coupling of the $Z$ to a
nucleus is expressed in terms of the {\it weak charge}, given approximately as
$Q_W \simeq \rho(N-Z-4Z \sin^2 \theta)$, where $N$ and $Z$ are the number of
neutrons and protons in the nucleus.  Very recently, a new result in atomic
cesium has been presented \cite{BW}: $Q_W = -72.06 \pm 0.28 \pm 0.35$, where
the first error is experimental and the second is theoretical. This result is
$2.5 \sigma$ from the theoretical prediction \cite{MR} of $Q_W = -73.20 \pm
0.13$  However, the deviation is opposite in sign from that caused by the the
most naive addition of particles in loops! This result bears watching.  The
experiment has been pushed about as far as it can go, so it is now incumbent
upon the theorists to check their calculations (and the refinements of them in
Ref.~\cite{BW} that reduced the theoretical error so dramatically from previous
values). 
\bigskip

\centerline{\bf V.  CP VIOLATION}
\bigskip

\leftline{\bf A.  The neutral kaon system}
\bigskip

The neutral kaon $K^0$ and its antiparticle $\bar K^0$ are an example of a
degenerate two-state system, with the degeneracy lifted by coupling to final
states.  So, too, are the two equal-frequency modes of a circular drum with a
single nodal line along the diameter.  Any basis may be chosen in which the
nodal line for one mode is perpendicular to that for the other. For example,
let the $K^0$ correspond to the mode with the node at 45 degrees with respect
to the $x$-axis; then the $\bar K^0$ will correspond to the orthogonal mode. 

Now a fly lands on the drum-head somewhere on the $x$ axis.  The two degenerate
states will be mixed and split in such a way that the fly couples to one mode
(with the node perpendicular to the $x$-axis) and not the other (with the node
along the $x$-axis).  The fly is like the $\pi \pi$ final state, and the
eigenstates are 
\beq
K_1 = \frac{K^0 + \bar K^0}{\st} ( \to \pi \pi)~~,~~~
K_2 = \frac{K^0 - \bar K^0}{\st} ( \not{\!\to} \pi \pi)~~~.
\eeq
Since the $\pi \pi$ system in the decay of the spinless kaons has even $CP$,
where $C$ is charge-reversal and $P$ is parity, or space inversion, the states
with definite mass and lifetime in the limit of CP conservation are $K_1$ and
$K_2$. The $K_1$ is thus much shorter-lived than the $K_2$, which has to decay
in some other manner than $\pi \pi$ \cite{GP}. 

In 1964, Christenson, Cronin, Fitch, and Turlay \cite{CCFT} discovered that the
long-lived kaon {\it also} decays to $\pi \pi$.  One could thus represent the
states of definite mass and lifetime as 
\beq \label{eqn:mix}
K_S \simeq K_1 + \epsilon K_2~~,~~~K_L \simeq K_2 + \epsilon K_1~~~.
\eeq
The parameter $\epsilon$ has a magnitude of a bit over $2 \times 10^{-3}$ and a
phase of about $\pi/4$.  Where does it come from? 

One possibility was suggested right after the discovery of CP violation:  A new
``superweak'' interaction \cite{SW} directly mixes $K^0$ and $\bar K^0$, with a
phase which leads to CP violation.  However, Kobayashi and Maskawa \cite{KM}
proposed that phases in the weak couplings of quarks to $W$ bosons generate
$\epsilon$ through loop graphs.  Three quark families are needed for
non-trivial phases.  The Kobayashi-Maskawa proposal thus entailed the existence
of the top and bottom quarks, later discovered at Fermilab. 

The loop graphs in question are ones in which, for example, a $K^0 = d \bar s$
undergoes a virtual transition via $W$ exchange to a pair $q_i \bar q_j$, where
$q_i$ and $q_j$ are any charge-2/3 quark:  $u,c,t$.  The $q_i \bar q_j$ pair
can then exchange a $W$ of the opposite charge to become $\bar K^0 = s \bar d$.
The top quark provides the dominant contribution to this process because of its
large mass. 

The Kobayashi-Maskawa (KM) theory of CP violation has recently survived two key
tests, the most recent of which seems to have firmly buried the superweak
theory.  These are results which Hiroshi would have enjoyed.  
\bigskip

\leftline{\bf B.  CP violation in $B$ meson decays}
\bigskip

The first new result concerns CP violation in the system of neutral $B$ mesons,
predicted to be large in the KM theory.  The same loop diagrams which mix
neutral kaons also mix $B^0 = d \bar b$ and $\bar B^0 = b \bar d$.  The phase
of the mixing amplitude is predicted within rather narrow limits by fits to
various weak-decay and mixing data. 

The best sign of CP violation in the $B$ meson system was anticipated
\cite{Sanda} to be the following asymmetry in rates:
\beq
A(J/\psi K_S) \equiv \frac{\Gamma(\bar B^0|_{t=0} \to J/\psi K_S)
 - \Gamma(B^0|_{t=0} \to J/\psi K_S)}
{\Gamma(\bar B^0|_{t=0} \to J/\psi K_S)
+ \Gamma(B^0|_{t=0} \to J/\psi K_S)} \ne 0~~~.
\eeq
Here the subscript indicates that the flavor of the neutral $B$ is identified
at the time of its production; it oscillates between $B^0$ and $\bar B^0$
thereafter as a result of $B^0$--$\bar B^0$ mixing.  The asymmetry arises from
the interference of the mixing amplitude with the decay amplitude.  The decay
$B^0 \to J/\psi K_S$ can occur either directly or through the sequential
process $B^0 \to \bar B^0 \to J/\psi K_S$, which imposes a modulating amplitude
on the direct decay.  The sign of this modulating amplitude is opposite to that
in $\bar B^0 \to B^0 \to J/\psi K_S$ (interfering with the direct $\bar B^0 \to
J/\psi K_S$ process), and so a difference arises in both time-dependent and
time-integrated rates. 

A recent result from the CDF Collaboration at Fermilab \cite{CDFbeta} observes
the asymmetry at about the $2 \sigma$ level with the value predicted by the KM
theory. Both SLAC and KEK are constructing ``$B$-factories'' to observe this
asymmetry (and many others) at a compelling statistical level, and many other
experiments (e.g., at Cornell, LEP, DESY and Fermilab) may have something to
say soon on CP-violation in $B$ decays.
\bigskip

\leftline{\bf C.  Demise of the superweak theory}
\bigskip

The second new result concerns the most significant result on the decays of
neutral kaons since the discovery that they violated CP in 1964.  Since then,
all CP-violating effects in the neutral kaon system could be parametrized by
the single quantity $\epsilon$ in Eq.~(\ref{eqn:mix}).  If that were so, one
should see no difference between the CP-violating decays $K_L \to \pi^+ \pi^-$
and $K_L \to \pi^0 \pi^0$ when normalized by the corresponding $K_S$ rates. 
Thus, the double ratio 
\beq
R \equiv \frac{\Gamma(K_L \to \pi^+ \pi^-)/\Gamma(K_S \to \pi^+ \pi^-)}
{\Gamma(K_L \to \pi^0 \pi^0)/\Gamma(K_S \to \pi^0 \pi^0)}
\eeq
should equal 1.  In the KM theory it can differ from 1 by up to a percent.
A ``direct'' decay amplitude, parametrized by a quantity $\epsilon'$, can
violate CP.  The double ratio is $R = 1 + 6 {\rm~Re}(\epsilon'/\epsilon)$.
The superweak theory has no provision for $\epsilon'$.

Two previous experiments gave conflicting results on whether $\epsilon'$ was
nonzero: 
\protect
\beq
{\rm~Re}(\epsilon'/\epsilon) = (7.4 \pm 5.9) \times 10^{-4}~~~
({\rm Fermilab~E731})~~~\cite{E731}~~~,
\eeq
\beq
{\rm~Re}(\epsilon'/\epsilon) = (23 \pm 6.5) \times 10^{-4}~~~
({\rm CERN~NA31})~~~\cite{NA31}~~~.
\eeq
A new experiment at Fermilab has now confirmed the CERN result with far more
compelling statistics, finding 
\beq \label{eqn:832}
{\rm~Re}(\epsilon'/\epsilon) = (28.0 \pm 4.1) \times 10^{-4}~~~
({\rm Fermilab~E832})~~~\cite{E832}~~~.
\eeq
The superweak theory is definitively ruled out.  The magnitude of the effect is
on the high end of the most recent theoretical range \cite{Buras}, but this may
merely represent a shortcoming of methods to estimate hadronic matrix elements
rather than any intrinsic limitation of the KM theory.  The new result will
probably reduce the uncertainty on the parameters of the CKM matrix. 
\bigskip

\leftline{\bf D.  Alternative sources of CP violation}
\bigskip

So far the Kobayashi-Maskawa theory of CP violation has survived experimental
tests.  But what if it is eventually ruled inconsistent or incomplete?  Many
other theories are lurking in the wings, including superweak {\it
contributions} to CP violation (clearly not the whole story), effects of
right-handed $W$ bosons, and multi-Higgs models.  These can be tested by a host
of forthcoming experiments, including those on rare kaon and $B$ meson decays,
searches for transverse muon polarization in the decays $K \to \pi \mu \nu$,
searches for neutron and electron electric dipole moments, and searches for CP
violation in decays of hyperons and charmed particles.  The field is very rich
and full of experimental opportunities. 
\bigskip

\centerline{\bf VI.  COMPOSITE HIGGS BOSONS}
\bigskip 

The SU(2) $\times$ U(1) electroweak gauge theory must be supplemented by a
mechanism for breaking the symmetry.  The standard (``Higgs'') mechanism
involves the introduction of an SU(2) doublet of complex scalar fields, or four
scalar mesons.  Three of the four scalars become the longitudinal components of
the $W$ and $Z$, and one remains as the physical Higgs boson.  With more than
one doublet, there will be additional observable scalar fields in the spectrum.

The Higgs fields interact with one another quartically in the Lagrangian. In
the presence of any physics beyond the electroweak scale, such as arises in
``grand unifications'' of the electroweak and strong interactions, such a
theory cannot be fundamental.  New physics must enter at a mass scale of a TeV
or less in order that the Higgs boson mass not receive large radiative
corrections from the higher mass scale.  Independently of grand unified
theories, the quartic Higgs interaction itself has undesirable high-energy
behavior, so that the only theory which makes sense is the ``trivial'' one in
which the quartic interaction vanishes. 

One approach to this problem is provided by supersymmetry, which provides a set
of ``superpartners'' to the currently observed particles, differing from them
by half a unit of spin.  The quartic interaction is then not fundamental, and
the superpartners cancel the large radiative corrections.  Another approach is
to postulate that the Higgs fields themselves are composite.  This idea
\cite{TC}, known as ``technicolor,'' envisions the Higgs fields as
fermion-antifermion pairs, with the new fermions bound by some new superstrong
force, in the same way that pions are made of quarks bound by the force of
quantum chromodynamics (QCD). In analogy with QCD, which implies a low-energy
pion-pion quartic interaction which is not really fundamental, the Higgs boson
quartic potential is then just a consequence of some more fundamental
underlying theory. 

Properties of the new technifermions can be learned by an argument based on
particles in loops.  Their charges must be such as to ensure anomaly
cancellation in the decay of a longitudinal $Z$ to two photons. If one has a
single SU(2) doublet $(U,D)$ of technifermions (occurring in some number of
``techni''-colors), the vanishing of $Q(U)^2 - Q(D)^2$ then requires $Q(U) =
1/2$, $Q(D) = -1/2$.  This was the original solution of ``minimal technicolor''
\cite{TC,Tera}. It was abandoned because there seems to be no evidence for
fundamental fermions with charges $\pm 1/2$, and because the minimal model only
explains the masses of $W$ and $Z$, not of quarks and leptons.  Attempts to
``extend'' technicolor to a theory of quark and lepton masses \cite{ETC}
introduce many new particles in loops and thus run afoul of the constraints
from precise electroweak experiments mentioned in Sec.~IV \cite{PT}.  In the
next section I will propose a solution \cite{JRCOMP} to which Hiroshi might
have been sympathetic, in view of his early efforts \cite{Vple} to uncover the
substructure of particles. 
\bigskip

\centerline{\bf VII.  COMPOSITE FERMIONS}
\bigskip

Suppose the minimal techniquarks $U$ and $D$ of Sec.~VII are the carriers of
the weak isospin (the SU(2) quantum number) in quarks and leptons.  A formula
for the charge of quarks and leptons which suggests this identification is
\cite{chg} $Q = I_{3L} + I_{3R} + (B-L)/2$, where $I_{3L}$ is left-handed
isospin, $I_{3R}$ is right-handed isospin, $B$ is baryon number, and $L$ is
lepton number.  We imagine $U,D$ to carry $I_{3L}$ and $I_{3R}$ since these
quantum numbers are naturally correlated with quark or lepton spin. (Note that
$I_{3L} + I_{3R}$ is always equal to $+1/2$ for up quarks and neutrinos and
$-1/2$ for down quarks and charged leptons.)  The $(B-L)/2$ contribution to the
charge then has to be carried by ``something else''.  Let it be a scalar $\bar
S_q$ with charge $1/6$ for three colors of quarks or $\bar S_\ell$ with charge
$-1/2$ for leptons.  The scalars thus belong to an SU(4)$_{\rm color}$ group
first proposed by Pati and Salam \cite{PaS}.  A $u$ quark is then $U \bar S_q$,
while an electron is $D \bar S_\ell$.  Tests of this model (or of others of
quark and lepton substructure) are possible at the highest LEP energies,
forthcoming Tevatron experiments, and future hadron and lepton colliders. 
\bigskip

\centerline{\bf VIII.  SUMMARY}
\bigskip

When this talk was originally given nearly five years ago at Minnesota, the top
quark had just been discovered, confirming a prediction based on its role in
loop diagrams.  Since then there have been great strides in confirming other
predictions of loop diagrams, including hints of CP violation in $B$ meson
decays and the overthrow of the superweak theory of CP violation. Experiments
in atomic parity violation suggest that we may not know the full story of
effects of particles in loops, but the presence of at least one puzzling result
is what makes our field interesting.  Hiroshi would have enjoyed the recent
developments.  On this occasion I extend good wishes to Akiko and to his
colleagues, and thank them for the opportunity to honor his memory. 
\bigskip

\centerline{\bf ACKNOWLEDGEMENTS}
\bigskip

I wish to thank S. Pakvasa for helpful comments on the early theory of charm.
This work was supported in part by the United States Department of Energy
under Contract No.~DE FG02 90ER40560.

\bigskip

\def \ajp#1#2#3{Am.~J.~Phys.~{\bf#1}, #2 (#3)}
\def \apny#1#2#3{Ann.~Phys.~(N.Y.) {\bf#1}, #2 (#3)}
\def \app#1#2#3{Acta Phys.~Polonica {\bf#1}, #2 (#3)}
\def \arnps#1#2#3{Ann.~Rev.~Nucl.~Part.~Sci.~{\bf#1}, #2 (#3)}
\def \art{and references therein}
\def \cmts#1#2#3{Comments on Nucl.~Part.~Phys.~{\bf#1}, #2 (#3)}
\def \cn{Collaboration}
\def \cp89{{\it CP Violation,} edited by C. Jarlskog (World Scientific,
Singapore, 1989)}
\def \dpfa{{\it The Albuquerque Meeting: DPF 94} (Division of Particles and
Fields Meeting, American Physical Society, Albuquerque, NM, Aug.~2--6, 1994),
ed. by S. Seidel (World Scientific, River Edge, NJ, 1995)}
\def \dpff{{\it The Fermilab Meeting: DPF 92} (Division of Particles and Fields
Meeting, American Physical Society, Batavia, IL., Nov.~11--14, 1992), ed. by
C. H. Albright \ite~(World Scientific, Singapore, 1993)}
\def \efi{Enrico Fermi Institute Report No. EFI}
\def \epl#1#2#3{Europhys.~Lett.~{\bf #1}, #2 (#3)}
\def \flg{{\it Proceedings of the 1979 International Symposium on Lepton and
Photon Interactions at High Energies,} Fermilab, August 23-29, 1979, ed. by
T. B. W. Kirk and H. D. I. Abarbanel (Fermi National Accelerator Laboratory,
Batavia, IL, 1979}
\def \hb87{{\it Proceeding of the 1987 International Symposium on Lepton and
Photon Interactions at High Energies,} Hamburg, 1987, ed. by W. Bartel
and R. R\"uckl (Nucl.~Phys.~B, Proc. Suppl., vol. 3) (North-Holland,
Amsterdam, 1988)}
\def \hpa#1#2#3{Helv.~Phys.~Acta {\bf#1}, #2 (#3)}
\def \ib{{\it ibid.}~}
\def \ibj#1#2#3{{\it ibid.}~{\bf#1}, #2 (#3)}
\def \ichep72{{\it Proceedings of the XVI International Conference on High
Energy Physics}, Chicago and Batavia, Illinois, Sept. 6 -- 13, 1972,
edited by J. D. Jackson, A. Roberts, and R. Donaldson (Fermilab, Batavia,
IL, 1972)}
\def \ijmpa#1#2#3{Int.~J. Mod.~Phys.~A {\bf#1}, #2 (#3)}
\def \ite{{\it et al.}}
\def \jpb#1#2#3{J.~Phys.~B~{\bf#1}, #2 (#3)}
\def \ky85{{\it Proceedings of the International Symposium on Lepton and
Photon Interactions at High Energy,} Kyoto, Aug.~19-24, 1985, edited by M.
Konuma and K. Takahashi (Kyoto Univ., Kyoto, 1985)}
\def \lkl87{{\it Selected Topics in Electroweak Interactions} (Proceedings of
the Second Lake Louise Institute on New Frontiers in Particle Physics, 15 --
21 February, 1987), edited by J. M. Cameron \ite~(World Scientific, Singapore,
1987)}
\def \lnc#1#2#3{Lettere al Nuovo Cim.~{\bf#1}, #2 (#3)}
\def \mpla#1#2#3{Mod.~Phys.~Lett.~A {\bf#1}, #2 (#3)}
\def \nc#1#2#3{Nuovo Cim.~{\bf#1}, #2 (#3)}
\def \np#1#2#3{Nucl.~Phys.~{\bf#1}, #2 (#3)}
\def \pisma#1#2#3#4{Pis'ma Zh.~Eksp.~Teor.~Fiz.~{\bf#1}, #2 (#3) [JETP Lett.
{\bf#1}, #4 (#3)]}
\def \pl#1#2#3{Phys.~Lett.~{\bf#1}, #2 (#3)}
\def \pla#1#2#3{Phys.~Lett.~A {\bf#1}, #2 (#3)}
\def \plb#1#2#3{Phys.~Lett.~{\bf#1}, #2 (#3)}
\def \pr#1#2#3{Phys.~Rev.~{\bf#1}, #2 (#3)}
\def \prc#1#2#3{Phys.~Rev.~C {\bf#1}, #2 (#3)}
\def \prd#1#2#3{Phys.~Rev.~D {\bf#1}, #2 (#3)}
\def \prl#1#2#3{Phys.~Rev.~Lett.~{\bf#1}, #2 (#3)}
\def \prp#1#2#3{Phys.~Rep.~{\bf#1}, #2 (#3)}
\def \ptp#1#2#3{Prog.~Theor.~Phys.~{\bf#1}, #2 (#3)}
\def \ptwaw{Plenary talk, XXVIII International Conference on High Energy
Physics, Warsaw, July 25--31, 1996}
\def \rmp#1#2#3{Rev.~Mod.~Phys.~{\bf#1}, #2 (#3)}
\def \rp#1{~~~~~\ldots\ldots{\rm rp~}{#1}~~~~~}
\def \rpp#1#2#3{Rep.~Prog.~Phys.~{\bf#1}, #2 (#3)}
\def \si90{25th International Conference on High Energy Physics, Singapore,
Aug. 2-8, 1990}
\def \slc87{{\it Proceedings of the Salt Lake City Meeting} (Division of
Particles and Fields, American Physical Society, Salt Lake City, Utah, 1987),
ed. by C. DeTar and J. S. Ball (World Scientific, Singapore, 1987)}
\def \slac89{{\it Proceedings of the XIVth International Symposium on
Lepton and Photon Interactions,} Stanford, California, 1989, edited by M.
Riordan (World Scientific, Singapore, 1990)}
\def \smass82{{\it Proceedings of the 1982 DPF Summer Study on Elementary
Particle Physics and Future Facilities}, Snowmass, Colorado, edited by R.
Donaldson, R. Gustafson, and F. Paige (World Scientific, Singapore, 1982)}
\def \smass90{{\it Research Directions for the Decade} (Proceedings of the
1990 Summer Study on High Energy Physics, June 25--July 13, Snowmass, Colorado),
edited by E. L. Berger (World Scientific, Singapore, 1992)}
\def \tasi90{{\it Testing the Standard Model} (Proceedings of the 1990
Theoretical Advanced Study Institute in Elementary Particle Physics, Boulder,
Colorado, 3--27 June, 1990), edited by M. Cveti\v{c} and P. Langacker
(World Scientific, Singapore, 1991)}
\def \waw{XXVIII International Conference on High Energy
Physics, Warsaw, July 25--31, 1996}
\def \yaf#1#2#3#4{Yad.~Fiz.~{\bf#1}, #2 (#3) [Sov.~J. Nucl.~Phys.~{\bf #1},
#4 (#3)]}
\def \zhetf#1#2#3#4#5#6{Zh.~Eksp.~Teor.~Fiz.~{\bf #1}, #2 (#3) [Sov.~Phys.~--
JETP {\bf #4}, #5 (#6)]}
\def \zpc#1#2#3{Zeit.~Phys.~C {\bf#1}, #2 (#3)}
\def \zpd#1#2#3{Zeit.~Phys.~D {\bf#1}, #2 (#3)}


\begin{thebibliography}{99}

\bibitem{Z3} J. L. Rosner, \prl{17}{1190}{1966}, \apny{44}{11}{1967}.

\bibitem{RS} J. L. Rosner and H. Suura, \pr{187}{1905}{1969}.

\bibitem{IR} H. Suura, \pr{99}{1020}{1955}; D. R. Yennie and H. Suura,
\pr{105}{1378}{1957}; H. Suura, \ptp{24}{225(L)}{1960}; D. R. Yennie, S. C.
Frautschi, and H. Suura, \apny{13}{379}{1961}; H. Suura and D. R. Yennie,
\prl{10}{69}{1963}.

\bibitem{SWP} H. Suura and E. H. Wichmann, \pr{105}{1930(L)}{1957}; A.
Petermann, \ibj{105}{1931(L)}{1957}. 

\bibitem{firstmu} For a review of early experiments see F. Combley and E.
Picasso, \prp{14}{1}{1974}. 

\bibitem{BNLgmu} C. Timmermans, presented at International Conference on High
Energy Physics, Vancouver, 23--29 July, 1998.

\bibitem{cmts} J. L. Rosner, \cmts{22}{205}{1998}.

\bibitem{FR} V. L. Fitch and J. L. Rosner, ``Elementary Particle Physics
in the Second Half of the Twentieth Century,'' Ch.~9 in {\it Twentieth
Century Physics}, edited by L. M. Brown, A. Pais, and B. Pippard (AIP/IOP,
New York and Bristol, 1995), pp.~635--794.

\bibitem{Schw} For the history of this and similar results see S. S. Schweber,
{\it QED and the Men Who Made It:  Dyson, Feynman, Schwinger, and Tomonaga}
(Princeton University Press, 1994).

\bibitem{SU} R. Serber, \pr{48}{49}{1935}; E. A. Uehling, \ibj{48}{55}{1935}.

\bibitem{Lamb} W. Lamb and R. C. Retherford, \pr{72}{241}{1947}.

\bibitem{Lambth} H. A. Bethe, \pr{72}{339}{1947}; N. M. Kroll and W. E. Lamb,
\ibj{75}{388}{1949}; J. Schwinger, \ibj{75}{898}{1949}; J. B. French and V. F.
Weisskopf, \ibj{75}{388(A), 1240}{1949}; R. P. Feynman, \ibj{76}{769}{1949};
Y. Nambu, \ptp{4}{82}{1949}.

\bibitem{Palch} V. G. Palchikov, L. Sokolov, and V. P. Yakovlev, Lett.~J.
Tech.~Phys.~{\bf38}, 347 (1983).

\bibitem{LP} S. R. Lundeen and F. M. Pipkin, \prl{46}{232}{1981}; Metrologia
{\bf 22}, 9 (1986). 

\bibitem{HP} E. W. Hagley and F. M. Pipkin, \prl{72}{1172}{1994}.

\bibitem{Pach} K. Pachucki, \prl{72}{3154}{1994}.

\bibitem{Kin} T. Kinoshita and D. R. Yennie, in {\it Quantum Electrodynamics},
edited by T. Kinoshita (Singapore, World Scientific, 1990), ch.~1.

\bibitem{Kinnew} T. Kinoshita, \rpp{59}{1459}{1996}.

\bibitem{Schwg} J. Schwinger, \pr{73}{416}{1948}, erratum \ibj{76}{790}{1949}.

\bibitem{Dehmelt} R. S. Van Dyck, Jr., P. B. Schwinberg, and H. G. Dehmelt,
\prl{59}{26}{1987}.  A new result by this group is anticipated:  See
\cite{Kinnew}. 

\bibitem{gmuth} T. Kinoshita and W. J. Marciano, in {\it Quantum
Electrodynamics} \cite{Kin}, ch.~10.

\bibitem{Cz} A. Czarnecki and M. Skrzypek, Brookhaven National Laboratory
report BNL-HET-98/38 (hep-ph/9812394) (unpublished), \art.

\bibitem{Davier} M. Davier, Orsay report LAL-98-87, hep-ph/9812370
(unpublished), \art.

\bibitem{gmuex} J. Bailey \ite, \pl{68B}{191}{1977}; F. J. M. Farley and E.
Picasso, in {\it Quantum Electrodynamics} (Ref.~\cite{Kin}), ch.~11. 

\bibitem{JS} J. Steinberger, \pr{76}{1180}{1949}; S. L. Adler, \pr{177}{2426}
{1969}; S. L. Adler and W. A. Bardeen, \ibj{182}{1517}{1969}; J. S. Bell and R.
Jackiw, \nc{60A}{47}{1969}. 

\bibitem{GZ} M. Gell-Mann, \pl{8}{214}{1964}; G. Zweig, CERN reports 8182/TH
401 (1964) and 8419/TH 412 (1964), unpublished; second paper reprinted in {\it
Developments in the Quark Theory of Hadrons} edited by D. B. Lichtenberg and S.
P. Rosen (Hadronic Press, Nonantum, Mass:, 1980), Press) v.~1, p.~22. 

\bibitem{HN} M. Y. Han and Y. Nambu, \pr{139}{B1006}{1965}.

\bibitem{Okubo} S. Okubo, in {\it Symmetries and Quark Models}, Proceedings of
the International Conference on Symmetries and Quark Models, Wayne State
University, 18--20 June 1969, edited by R. Chand (Gordon and Breach, New York,
1970), p.~59.

\bibitem{OS} J. Otokozawa and H. Suura, \prl{21}{1295}{1968}; H. Suura and
B.-L. Young, \nc{11A}{101}{1972}; B.-L. Young, H. Suura, and T. F. Walsh,
\lnc{4}{505}{1972}. 

\bibitem{Wilcz} M. Alford, K. Rajagopal, and F. Wilczek, \np{B537}{443}{1999};
T. Schafer and F. Wilczek, Institute for Advanced Study report IASSNS-98-100
(hep-ph/9811473), unpublished; J. Berges, MIT report MIT-CTP-2829
(hep-ph/9902419), unpublished, \art. 

\bibitem{BIM} C. Bouchiat, J. Iliopoulos, and Ph.~Meyer,
\pl{38B}{519--23}{1972}; H. Georgi and S.L. Glashow, \prd{6}{429}{1972}; D.
J. Gross and R. Jackiw, \ibj{6}{477}{1972}. 

\bibitem{charm} J. J. Aubert \ite, \prl{33}{1404}{1974}; J.-E. Augustin \ite,
\ibj{33}{1406}{1974}.

\bibitem{Niu} K. Niu, E. Mikumo, and Y. Maeda, \ptp{46}{1644}{1971}.

\bibitem{charmth} T. Hayashi \ite, \ptp{47}{280, 1998}{1972}; \ibj{49}{350,
353}{1973}; \ibj{52}{636}{1974}. 

\bibitem{Perl} M. L. Perl \ite, \prl{35}{1489}{1975}; \pl{63B}{466}
{1976}; \ibj{70B}{487}{1977}.

\bibitem{KM} M. Kobayashi and T. Maskawa, \ptp{49}{652}{1973}.

\bibitem{HH} H. Harari, in {\it Proc. 1975 Int. Symp. on Lepton and Photon
Interactions (Stanford University, August 21--27, 1975)}, edited by W. T. Kirk
(Stanford Linear Accelerator Center, Stanford, CA, 1976), p.~317; H.
Harari, in Proceedings of the 20th Annual SLAC Summer Institute on Particle
Physics: {\it The Third Family and the Physics of Flavor}, edited by L.
Vassilian, Stanford Linear Accelerator Center report SLAC-412, p.~647.

\bibitem{ups} S. W. Herb \ite, \prl{39}{252}{1977}; W. R. Innes \ite, \ibj{39}
{1240, 1640(E)}{1977}.

\bibitem{top} CDF \cn, F. Abe \ite, \prl{73}{225}{1994}, \prd{50}{3966}{1994};
\ibj{51}{4623}{1995}; \prl{74}{2626}{1995}; D0 \cn, S. Abachi \ite, \ibj{74}
{2632}{1995}; \prd{52}{4877}{1995}.

\bibitem{topmass} R. Partridge, Rapporteur's Talk, presented at International
Conference on High Energy Physics, Vancouver, 23--29 July, 1998.

\bibitem{GWS} S. L. Glashow, \np{22}{579}{1961}; S. Weinberg,
\prl{19}{1264}{1967}; A. Salam, in {\it Proceedings of the Eighth Nobel
Symposium}, edited by N. Svartholm (Almqvist and Wiksell, Stockholm; Wiley, New
York, 1978), p. 367. 

\bibitem{Wdisc} UA1 \cn, G. Arnison \ite, \pl{122B}{103}{1983}; \ibj{129B}
{273}{1983}; UA2 \cn, M. Banner \ite, \pl{122B}{476}{1983}.

\bibitem{Zdisc} UA1 \cn, G. Arnison \ite, \pl{126B}{398}{1983}; UA2 \cn,
\pl{129B}{130}{1983}. 

\bibitem{NC} F. J. Hasert \ite, \pl{46B}{121,138}{1973}; \np{B73}{1}{1974};
A. Benvenuti \ite, \prl{32}{800}{1974}; B. Aubert \ite, \ibj{32}{1454}{1974}.

\bibitem{Cab} N. Cabibbo, \prl{10}{531}{1963}.

\bibitem{Karlen} D. Karlen, Rapporteur's Talk, presented at International
Conference on High Energy Physics, Vancouver, 23--29 July, 1998.

\bibitem{Veltman} M. Veltman, \np{B123}{89}{1977}.

\bibitem{LQT} See, e.g., M. Veltman, \app{B8}{475}{1977} \pl{70B}{253}{1977};
B. W. Lee, C. Quigg, and H. B. Thacker, \prl{38}{883}{1977};
\prd{16}{1519}{1977}; C. Quigg, {\it Gauge Theories of the Weak,
Electromagnetic, and Strong Interactions}, Benjamin/Cummings, 1983. 

\bibitem{PT} M. Peskin and T. Takeuchi, \prl{65}{964}{1990}; \prd{46}{381}
{1992}.

\bibitem{mod} This result is slightly altered to $M_H = (105^{+73}_{-46})
{\rm~GeV}/c^2$ by the new value of $\alpha(M_Z)$ reported in \cite{Davier}.

\bibitem{MR} W. Marciano and J. L. Rosner, \prl{65}{2963}{1990}; \ibj{68}
{898(E)}{1992}.

\bibitem{Sandars} P. G. H. Sandars, \jpb{23}{L655}{1990}.

\bibitem{BW} S. C. Bennett and C. E. Wieman, University of Colorado report,
January, 1999, to be published in Phys. Rev. Letters.

\bibitem{GP} M. Gell-Mann and A. Pais, \pr{97}{1387}{1955}; T. D. Lee, R.
Oehme, and C. N. Yang, \pr{106}{340}{1957}; B. L. Ioffe, L. B. Okun', and A. P.
Rudik, \zhetf{32}{396}{1957}{5}{328}{1957}. 

\bibitem{CCFT} J. H. Christenson, J. W. Cronin, V. L. Fitch, and R. Turlay,
\prl{13}{138}{1964}.

\bibitem{SW} L. Wolfenstein, \prl{13}{562}{1964}.

\bibitem{Sanda} I. I. Bigi and A. I. Sanda, \np{B193}{85}{1981}.

\bibitem{CDFbeta} CDF \cn, report CDF/PUB/BOTTOM/CDF/4855, preliminary version
released 5 February 1999.

\bibitem{E731} Fermilab E731 \cn, L. K. Gibbons \ite, \prl{70}{1203}{1993}.

\bibitem{NA31} CERN NA31 \cn, G. D. Barr \ite, \plb{317}{233}{1993}.

\bibitem{E832} Fermilab E832 (KTeV) \cn, presented by P. Shawhan at Fermilab,
24 February 1999.

\bibitem{Buras} A. J. Buras, M. Jamin, and M. E. Lautenbacher, \plb{389}{749}
{1996}.

\bibitem{TC} S. Weinberg, \prd{13}{974}{1976}; \ibj{19}{1277}{1979};
L. Susskind, \prd{20}{2619}{1979}.

\bibitem{Tera} H. Terazawa, \prd{22}{184}{1980}.

\bibitem{ETC} S. Dimopoulos and L. Susskind, \np{B155}{237}{1979}; E. Eichten
and K. Lane, \pl{90B}{237}{1980}; E. Eichten, I. Hinchliffe, K. D. Lane, and
C. Quigg, \prd{34}{1547}{1986}.

\bibitem{JRCOMP} For more details and references, see J. L. Rosner, \efi~98-60,
hep-ph/9812537, to be published in Comments on Nuclear and Particle Physics. 

\bibitem{Vple} H. Suura, \ptp{6}{893}{1951}; K. Higashijima, V. Vi\v{s}nji
\'{c}, and H. Suura, \prd{30}{655}{1984}; H. Suura, ``A C-Number Dynamical
Model for Fermion Generations,'' University of Minnesota report UMN-TH-839/90
(unpublished).

\bibitem{chg} J. C. Pati and A. Salam, \prd{10}{275}{1974}; G. Senjanovic and
R. N. Mohapatra, \prd{12}{1502}{1975}; A. Davidson, \prd{20} {776}{1979}; R.
Marshak and R. Mohapatra, \pl{91B}{222}{1980}. 

\bibitem{PaS} J. C. Pati and A. Salam, \prd{10}{275}{1974}; J. C. Pati, A.
Salam, and J. Strathdee, \pl{59B}{265}{1975}. 

\bibitem{GS} O. W. Greenberg and J. Sucher, \pl{99B}{339}{1981}.  For
fermion-scalar models of composite fermions see also
O. W. Greenberg, \prl{35}{1120}{1975}; 
M. Veltman, in \flg, p.~529;
H. Fritzsch and G. Mandelbaum, \pl{102B}{319}{1981};
R. Casalbuoni and R. Gatto, \pl{103B}{113}{1981};
O. W. Greenberg, R. N. Mohapatra, and S. Nussinov, \pl{148B}{465}{1984};
M. Suzuki, \prd{45}{1744}{1992}.

\end{thebibliography}
\end{document}